\documentclass[aps,showpacs,superscriptaddress,twocolumn]{revtex4}%
\usepackage{amsfonts}
\usepackage{amsmath}
\usepackage{amssymb}
\usepackage{graphicx}%
\begin{document}
\title{Fractional topological excitations and quantum phase transition in a bilayer
2DEG adjacent to a superconductor film }
\author{Ningning Hao}
\affiliation{Institute of Applied Physics and Computational
Mathematics, P.O. Box 8009, Beijing 100088, People's Republic of
China} \affiliation{Institute of Physics, Chinese Academy of
Sciences, Beijing 100080, People's Republic of China}
\author{Wei Zhang}
\affiliation{Institute of Applied Physics and Computational
Mathematics, P.O. Box 8009, Beijing 100088, People's Republic of
China}
\author{Zhigang Wang}
\affiliation{Institute of Applied Physics and Computational
Mathematics, P.O. Box 8009, Beijing 100088, People's Republic of
China}
\author{Yupeng Wang}
\affiliation{Institute of Physics, Chinese Academy of Sciences, Beijing 100080, People's
Republic of China}
\author{Ping Zhang}
\thanks{Corresponding author. zhang\_ping@iapcm.ac.cn}
\affiliation{Institute of Applied Physics and Computational
Mathematics, P.O. Box 8009, Beijing 100088, People's Republic of
China} \affiliation{Center for Applied Physics and Technology,
Peking University, Beijing 100871, People's Republic of China}

\begin{abstract}
We study a bilayer two-dimension-electron-gas (2DEG) adjacent to a
type-II superconductor thin film with a pinned vortex lattice. We
find that with increasing interlayer tunneling, the system of half
filling presents three phases: gapped phase-I (topological
insulator), gapless critical phase-II (metal), and gapped phase-III
(band insulator). The Hall conductance for phase-I/III is 2/0
$e^{2}/h$, and has non-quantized values in phase-II. The excitation
(response to topological defect, a local vortex defect) in these
three phases shows different behaviors due to the topological
property of the system, including fractional charge $e/2$ for each
layer in phase-I. While in the case of quarter filling, the system
undergoes a quantum phase transition from metallic phase to
topological insulator phase (with excitation of fractional charge
$e/4$).

\end{abstract}
\maketitle

%\section
\textbf{Introduction} Topological excitation and fractional charge have
attracted considerable attention for decades. Jackiw and Rebbi first found
topological excitation with fractionalized charge in one-dimensional (1D)
system of a Dirac fermion field coupling with a topologically nontrivial Bose
field \cite{one}. Su $et$ $al.$ provided a nice intuitive picture for
charge\ $e/2$ soliton in polyacetylene chain \cite{two}. Goldstone and Wilczek
discussed the possibility of irrational charge in 1+1D systems \cite{wilczek}.
The two-dimensional (2D) fractional quantum Hall (QH) system with strong
correlation and time-reversal symmetry (TRS) broken, firstly studied by Tsui
$et$ $al$ \cite{three}, supports elementary excitations of the many-body
ground state with fractional charge and fractional statistics \cite{three1}.
Recently, Hou $et$ $al.$ \cite{four} have studied a model of graphene-like
structures with a vortex configuration in the Kekul\'{e} modulations of the
hopping amplitudes. They have found a fractional charge $e/2$ bonded to the
vortex without breaking TRS. Similar phenomenon was found in a model of 2D
square lattice with the analogical modulations of the hopping amplitudes
\cite{five}. These results indicate that the TRS and strong correlation are
not the necessary conditions for charge fractionalization in 2D.

Weeks $et$ $al.$ \cite{six} proposed another experimentally accessible 2D
weakly interacting system of two-dimensional electron gas (2DEG) in the
integer QH state adjacent to a film of type-$\Pi$ superconductor supplying the
quantization of flux in units of $\frac{1}{2}\Phi_{0}$ ($\Phi_{0}=h/e$). They
found the excitations with fractional charge and anyonic statistics which can
be described by a wave function composed by a set of filled one-particle
states. Moreover, the system could be fabricated in the laboratory \cite{six1}.

As is known, bilayer system may show new interesting features compared with
those of the corresponding monolayer system, for example, the bilayer QH
system versus the monolayer QH system. In this paper, we study the bilayer
2DEG adjacent to a film of type-$\Pi$ superconductor. The feasibility of the
similar system with monolayer 2DEG was argued by Seradjeh \textit{et al}. in
detail \cite{six1}. Our studies show that the system undergoes quantum phase
transitions from gapped phase-I/ topological insulator (TI) to gapless
critical phase-II/metal, then to gapped phase-III/band insulator (BI) with
increasing interlayer tunneling at half filling (here, half filling means one
electron per magnetic cell; note that a magnetic cell is twice of a monolayer
crystalline cell). These quantum phase transitions are not related to any
symmetry breaking. The difference between TI phase and BI phase is that the
energy spectrum of the edged system in the former has gapless topological edge
state. Furthermore, we find that the three phases have different Hall
conductances and excitations as response to the flux defect, i.e., an extra or
missing $\frac{1}{2}\Phi_{0}$ flux in the vortex lattices shows different
charge density profiles. The excitation of the BI phase has charge $e/2$ for
each layer. While at quarter filling case, there are one gapless metallic
phase and one gapped TI phase with excitation of charge $e/4$ for each layer.

%\section
\textbf{Lattice model and energy spectrum} The tight-binding Hamiltonian for
independent electrons in the presence of square vortex lattice is given by
\begin{align}
&  \hspace{-10pt}H_{latt}=-\underset{<i,j>\alpha}{\sum}t_{ij}e^{i\theta
_{ij\alpha}}c_{i\alpha}^{+}c_{j\alpha}-\underset{<<i,j>>\alpha}{\sum}%
t_{1ij}e^{i\varphi_{ij\alpha}}c_{i\alpha}^{+}c_{j\alpha}\nonumber\\
&  -t_{2}\underset{i,\alpha\neq\beta}{\sum}c_{i\alpha}^{+}c_{i\beta},
\end{align}
where $\alpha$($\beta$)$\mathtt{=}1$,$2$ refers to different layers,
$\mathtt{<}i,j\mathtt{>}$ $(\mathtt{<<}i,j\mathtt{>>})$\ represents
the nearest-neighbor (next-nearest-neighbor) sites, $t_{ij}$
($t_{1ij}$) is hopping amplitude between nearest-neighbor
(next-nearest-neighbor) sites in the same layer, and $t_{2}$ is the
interlayer tunneling amplitude.
%$V$ is a uniform on-site potential supported by the bias voltages.
In this work, for simplicity, we assume that $t_{ij}\mathtt{=}t$ and
$t_{1ij}\mathtt{=}t_{1}$. We have neglected the electron-electron interaction
and also assumed that the spins of all electrons are polarized along the
field. $c_{i\alpha}$ annihilates an electron at site $\mathbf{r}_{i}$ in layer
$\alpha.$ In the following we set the lattice constant $a$=1, $t$=1, $t_{1}%
/t$=$\gamma_{1}$, and $t_{2}/t$=$\gamma_{2}$.

The effect of magnetic field is included through the Peierls phase factors
\[
\ \ \ \ \theta_{ij\alpha}=\frac{2\pi}{\Phi_{0}}\underset{<i,j>\alpha}%
{\int_{\mathbf{r}_{i}}^{\mathbf{r}_{j}}}\mathbf{A}_{\alpha}\cdot
d\mathbf{r},\ \ \varphi_{ij\alpha}=\frac{2\pi}{\Phi_{0}}\underset
{<<i,j>>}{\int_{\mathbf{r}_{i}}^{\mathbf{r}_{j}}}\mathbf{A}_{\alpha}\cdot
d\mathbf{r},
\]
with $\mathbf{A}_{\alpha}$=$\frac{\Phi_{0}}{2}(0,x)$ the vector potential in
Landau gauge so that each plaquette in layer $\alpha$ is uniformly threaded by
a flux $\frac{1}{2}\Phi_{0}.$ The magnetic unit cell in layer $\alpha
$\ includes two sites $(l,m)\alpha$ and $(l\mathtt{+}1,m)\alpha$ denoted by
$A\alpha$ and $B\alpha$. In bilayer system, the wave function is of the form
of spinor field $\psi(\mathbf{r})$=$(c_{B1}(\mathbf{r}_{l+1m}),c_{A1}%
(\mathbf{r}_{lm}),c_{B2}(\mathbf{r}_{l+1m}),c_{A2}(\mathbf{r}_{lm}),)^{T}$.
The bilayer Hamiltonian can be diagonalized in the momentum space, with the
reduced Brillouin zone $BZ$=$\{\mathbf{k}:\left\vert k_{x}\right\vert
\mathtt{\leq}\pi/2,\left\vert k_{y}\right\vert \mathtt{\leq}\pi\},$ as
$H_{latt}\mathtt{=}\sum_{\mathbf{k}}\psi_{\mathbf{k}}^{+}\tilde{H}%
_{\mathbf{k}}\psi_{\mathbf{k}},$ where $\psi_{\mathbf{k}}\mathtt{=}%
(c_{B1}(\mathbf{k}),c_{A1}(\mathbf{k}),c_{B2}(\mathbf{k}),c_{A2}%
(\mathbf{k}))^{T}$. Furthermore, after a rotation of the spinor field,
$\varphi_{\mathbf{k}}\mathtt{=}S^{+}\psi_{\mathbf{k}}$, with $\ S\mathtt{=}%
\exp(i\pi I\mathtt{\otimes}\sigma_{x}/4)\exp(i\pi
 I\mathtt{\otimes}\sigma_{z}/4)\exp(i\pi
 I\mathtt{\otimes}\sigma_{x}/2),$ with $\sigma_{x,y,z}$
the Pauli matrices, the Hamiltonian $H_{\mathbf{k}}$ has a
simplified form
\begin{align}
H_{\mathbf{k}}  &  =S^{+}\tilde{H}S= I\otimes(2\cos k_{y}\sigma
_{x}+2\cos k_{x}\sigma_{y}\nonumber\\
&  +4\gamma_{1}\sin k_{x}\sin k_{y}\sigma_{z})+\gamma_{2}\sigma_{x}%
\otimes I.
\end{align}
We define the operator $G\mathtt{\equiv}i\sigma_{y}\mathtt{\otimes }
I$ and it is easy to see that $GH^{\ast}G\mathtt{=}H$,
$G^{2}\mathtt{=-}1$. As a consequence, for each eigenstate
$\psi_{E}$ with eigenvalue $E$, there is a corresponding eigenstate
$G\psi_{E}^{\ast}$ with eigenvalue $-E$. The energy bands of
$H_{\mathbf{k}}$ are given explicitly by
\begin{align}
E_{1}(\mathbf{k})  &  =\gamma_{2}+\varepsilon(\mathbf{k}),E_{2}(\mathbf{k}%
)=\gamma_{2}-\varepsilon(\mathbf{k}),\nonumber\\
E_{3}(\mathbf{k})  &  =-\gamma_{2}+\varepsilon(\mathbf{k}),E_{4}%
(\mathbf{k})=-\gamma_{2}-\varepsilon(\mathbf{k}),
\end{align}
where $\varepsilon(\mathbf{k})$=\texttt{ }2$\sqrt{\cos^{2}k_{x}+\cos^{2}%
k_{y}+4\gamma_{1}^{2}\sin^{2}k_{x}\sin^{2}k_{y}}$. The energy bands are
symmetric about zero energy. The bands are sketched in Fig. 1 for different
values of interlayer tunneling $\gamma_{2}$. There are four branches of the
curves corresponding to Eq. (4). For the half filling case with $0\mathtt{<}%
\gamma_{2}\mathtt{<}4\gamma_{1}$, the bands $E_{2}(\mathbf{k})$ and
$E_{3}(\mathbf{k})$ are separated with a gap $\Delta\mathtt{=}8\gamma
_{1}\mathtt{-}2\gamma_{2}$. The system is in a TI phase due to the existence
of gapless edge states (with topological winding numbers) for the edged sample
(the detailed calculation is not shown in this paper), or nontrivial Chern
number as seen later. When $4\gamma_{1}\mathtt{<}\gamma_{2}\mathtt{<}$
$2\sqrt{2}$ \cite{note1}, the bands $E_{2}(\mathbf{k})$ and $E_{3}%
(\mathbf{k})$ mix each other, leading to disappearance of the gap. For
$\gamma_{2}\mathtt{>}2\sqrt{2}$, the gap reappear, showing the behavior of a
usual BI. For the case of quarter filling, when $\gamma_{2}\mathtt{<}$
$\sqrt{2}\mathtt{-}2\gamma_{1}$, the bands $E_{3}$ and $E_{4}$ mix each other
leading to gapless metallic phase; when $\gamma_{2}\mathtt{>}$ $\sqrt
{2}\mathtt{-}2\gamma_{1}$, the system is always in the TI phase with a gap
$\Delta$=$2(\gamma_{2}\mathtt{-}(\sqrt{2}\mathtt{-}2\gamma_{1}))$.
\begin{figure}[ptb]
\begin{center}
\includegraphics[width=1\linewidth]{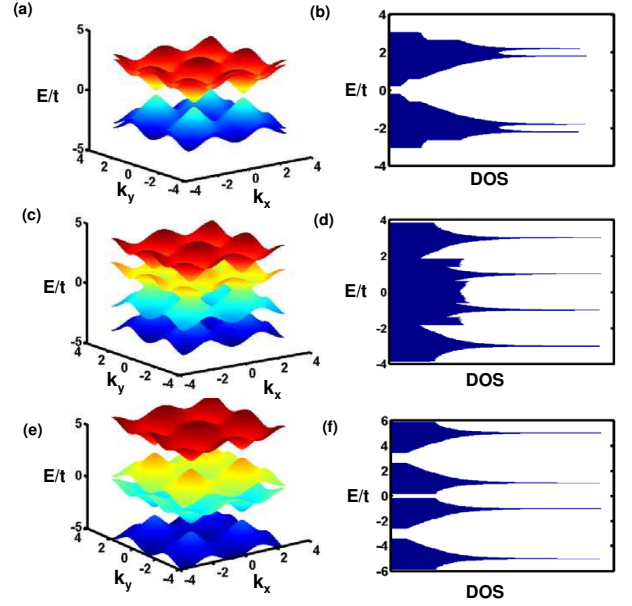}
\end{center}
\caption{(Color online) The energy band structures of the lattice Hamiltonian
(1) with different $\gamma_{2}:$ (a) $\gamma_{2}=0.2$, (c) $\gamma_{2}=1.0$,
(e) $\gamma_{2}=3.0$. $\gamma_{1}=0.1.$ The corresponding densities of the
states in arbitrary unit are shown in (b), (d) and (f). }%
\label{fig1}%
\end{figure}

%\section{QUANTUM PHASE TRANSITION}
\textbf{Quantum phase transitions}
%In order to illuminate the system undergoes the $QPT$ \ as
We look at the quantum phase transitions with changing of the interlayer
tunneling from further other point of view. First, we rewrite the system's
Hamiltonian with $\gamma_{2}$ as the control parameter,%

\begin{equation}
H_{\mathbf{k}}(\gamma_{2})=H_{\mathbf{k}}^{(0)}+\gamma_{2}H_{\mathbf{k}%
,}^{(1)}%
\end{equation}
where $H_{\mathbf{k}}^{(0)}$=$ I\mathtt{\otimes}(2\cos k_{y}%
\sigma_{x}\mathtt{+}2\cos k_{x}\sigma_{y}\mathtt{+}4\gamma_{1}\sin
k_{x}\sin k_{y}\sigma_{z})$ and
$H_{\mathbf{k}}^{(1)}\mathtt{=}\sigma_{x}\mathtt{\otimes } I$. Note
that
%$%\sigma _{y}H_{\mathbf{k}}^{(0)\ast }\sigma _{y}=-H_{\mathbf{k}}^{(0)}$, $%
%\sigma _{y}H_{\mathbf{k}}^{(1)\ast }\sigma _{y}=H_{\mathbf{k}}^{(1)}$ and
$[H_{\mathbf{k}}^{(0)},H_{\mathbf{k}}^{(1)}]\mathtt{=}0$. Thus quantum phase
transition may appear as a few lowest energy levels cross and the properties
of ground state change dramatically \cite{nine}. One may also find possible
quantum phase transitions by calculating the ground-state fidelity, defined as
the overlap between $\Psi_{0}(\gamma_{2})$ and $\Psi_{0}(\gamma_{2}%
\mathtt{+}\delta)$ \cite{eight},
\begin{equation}
F(\gamma_{2},\delta)=\left\vert \left\langle \Psi_{0}(\gamma_{2}+\delta
)|\Psi_{0}(\gamma_{2})\right\rangle \right\vert ,
\end{equation}
where $\delta$ is a small quantity, and the many-body ground-state wave
functions of the system $\Psi_{0}$ can be constructed with the one-particle
wave functions through the Slater determinant. In the numerical calculation of
the fidelity, we have used 40000 basis and $\delta$=0.005. The fidelity as a
function of $\gamma_{2}$ is shown in Fig. 2. We see that it is always constant
and equal to unity in the two regions of $0\mathtt{<}\gamma_{2}\mathtt{<}%
4\gamma_{1}$ and $\gamma_{2}\mathtt{>}$ $2\sqrt{2}$ for the half filling case,
indicating that each region corresponds to a specific phase. The fact that the
fidelity is equal to zero in the region $4\gamma_{1}\mathtt{<}\gamma
_{2}\mathtt{<}2\sqrt{2}$ tells us the system is in the critical phase in this
region, which is consistent with the absence of gap and lack of characteristic
length scale in phase-II. Similarly, in the case of quarter filling ($\nu
$=1/4), there are only two phases separated by the critical point
$\gamma_{2,c}$=\texttt{ }$\sqrt{2}\mathtt{-}2\gamma_{1}$. These results are
consistent with what we have found based on the analysis of the energy
spectrums and can be described by the energy-level crossing \cite{nine}.

%\section{TOPOLOGICAL EXCITATION}
\textbf{Topological excitations and factional charges } We first calculate the
Hall conductance based on the Kubo formula $\sigma_{H}$=$\frac{e^{2}}{h}C$
with $C$=$\frac{1}{2\pi}\underset{n}{\sum}\int^{E_{F}}d^{2}k\hat{z}%
\cdot(\nabla_{\mathbf{k}}\times A_{n}(\mathbf{k}))$, where the upper limit
means that the integration is over all occupied states below the Fermi energy
$E_{F}$. Here $A_{n}(\mathbf{k})$=$i\left\langle \psi_{n}(\mathbf{k}%
)\right\vert \left.  \nabla_{\mathbf{k}}\psi_{n}(\mathbf{k)}\right\rangle $ is
the Berry phase connection for $n$th band with wave function $\psi
_{n}(\mathbf{k)}$. When the Fermi level is in the gap, one has the well-known
(Thouless, Kohmoto, Nightingale, and Nijs) TKNN formula \cite{seven} $C$%
=$\sum_{n}C_{n}$, where $C_{n}=\frac{1}{2\pi}\int_{BZ}d^{2}k\hat{z}%
\cdot(\nabla_{\mathbf{k}}\times A_{n}(\mathbf{k}))$ is the Chern number. The
results for the Hall conductance is shown in Fig. 3, which reflect the
different properties of the three (two) phases for the case of half (quarter)
filling. It is clearly seen that in the half filling case, the Hall
conductance is 2/0 (in unit of $e^{2}/h$) in the phase-I/III, and is not
quantized in the gapless phase-II; in the quarter filling case, the Hall
conductance is 1 in the gapped phase-II, and is not quantized in the gapless phase-I.

The topological excitation appears as the response to the perturbation added
to the system. Here the perturbation is in a form of a flux defect $\eta
\Phi_{0}$ ($\eta$=$\pm\frac{1}{2}$) to the vortex lattice as discussed in Ref.
[8]. The vector potential in each layer has the following modulation
$\delta\mathbf{A}_{\alpha}$=$\frac{\eta\Phi_{0}}{2\pi r^{2}}\left(
\mathbf{r}\times\hat{z}\right)  $. Correspondingly, the modulation to Peierls
factors is
\begin{align}
\delta\theta_{ij\alpha}  &  =\frac{2\pi}{\Phi_{0}}\underset{<i,j>\alpha}%
{\int_{\mathbf{r}_{i}}^{\mathbf{r}_{j}}}\delta\mathbf{A}_{\alpha}\cdot
d\mathbf{r=}\eta\underset{<i,j>\alpha}{\int_{\mathbf{r}_{i}}^{\mathbf{r}_{j}}%
}d\theta\nonumber\\
\delta\varphi_{ij\alpha}  &  =\frac{2\pi}{\Phi_{0}}\underset{<<i,j>>\alpha
}{\int_{\mathbf{r}_{i}}^{\mathbf{r}_{j}}}\delta\mathbf{A}_{\alpha}\cdot
d\mathbf{r=}\eta\underset{<<i,j>>\alpha}{\int_{\mathbf{r}_{i}}^{\mathbf{r}%
_{j}}}d\theta.
\end{align}
In a stringy gauge, the Peierls factors can be specified as $\delta\theta
_{ij}=\pi$ ($\delta\varphi_{ij}=\pi$) if the string, originating from the
defect and ending at a boundary, cuts the $i$-$j$ bond, and zero otherwise.
These extra Peierls factors have the properties $\oint\delta\theta_{ij}$=$\pi$
mod $2\pi$ ($\oint\delta\varphi_{ij}$=$\pi$ mod $2\pi$) for closed loop
containing the defect. They look like a vortex profile. These topologically
nontrivial Peierls factors have profound effects on the behavior of the excitation.

\begin{figure}[ptb]
\begin{center}
\includegraphics[width=1\linewidth]{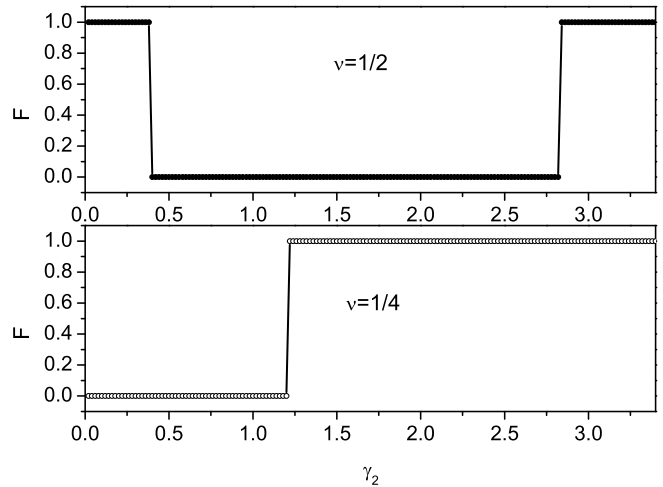}
\end{center}
\caption{The fidelity versus inter-layer coupling $\gamma_{2}$. $\gamma
_{1}=0.1$. }%
\label{fig2}%
\end{figure}

\begin{figure}[ptb]
\begin{center}
\includegraphics[width=1\linewidth]{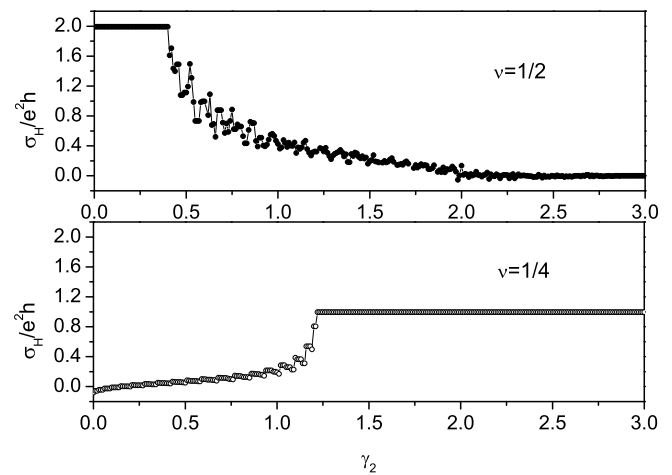}
\end{center}
\caption{The Hall conductance versus inter-layer coupling $\gamma_{2}$.
$\gamma_{1}=0.1$. }%
\label{fig3}%
\end{figure}

\begin{figure}[ptb]
\begin{center}
\includegraphics[width=1\linewidth]{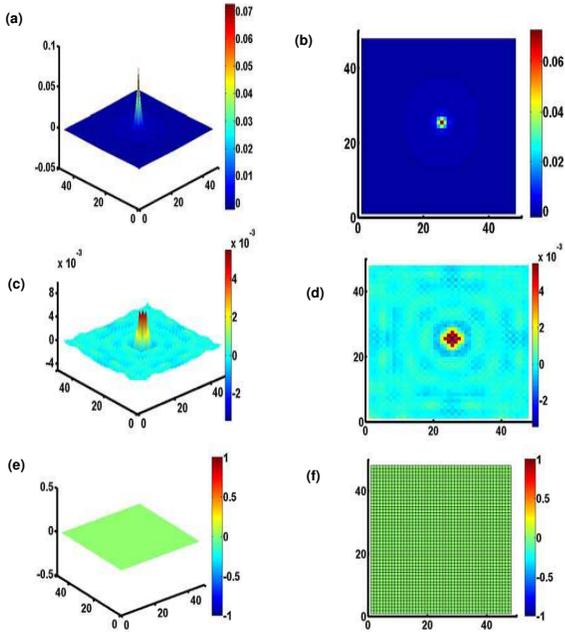}
\end{center}
\caption{(Color online) The charge densities of the excitation for different
$\gamma_{2}:$ (a) $\gamma_{2}=0.2$, (c) $\gamma_{2}=1.0$, (e) $\gamma_{2}%
=3.0$, in a $48\times48$ lattice system with periodic boundary condition. (b),
(d), (f) are the corresponding planforms. In all cases, we have subtracted the
background charge density which is uniformly equal to $0.5$ on each site of
the lattice at the half filling case. $\gamma_{1}$=$0.1.$}%
\label{fig4}%
\end{figure}

Let us look at the basic properties of the excitation, as the response of the
system to the flux defect, in the three phases for the case of half filling.
Since phase-I and-III are gapped, thus are incompressible, the charge for each
layer due to the presence of the flux defect is $\delta Q$=$\eta\sigma_{H}%
e/2$, as argued in \cite{six}. Combining with above results of the
conductance, we have $\delta Q$=$e/2$ (for $\eta$=$1/2$) in phase-I and
$\delta Q$=$0$ in phase-III. To confirm this result and gain more
understanding of the properties of the three phases, we perform a numerical
calculation based on exact diagonalizations of lattice Hamiltonian. We have
applied this method to the systems of sizes up to $48\times48$ at different
values of interlayer tunneling $\gamma_{2}$ and obtained the charge density
profile (per layer) as presented in Fig. 4. The two layers have the same
results since the defects in each layer are superposable along the magnetic
field direction. By integration of the density profile, we get the charge for
each layer $\delta Q$=$e/2$ (0) for the gapped phase-I (III), agreeing with
the above arguments based on the incompressibility and Hall conductance. The
situation is quite different for the gapless phase II. The charge profile is
nonlocal and there is no well-defined localized charge due to the lack of
characteristic length scale in the gapless critical phase-II. This is unlike
the gapped phase-I, where the presence of localized charge is due to the
nontrivial Peierls phase profile and characteristic length scale determined by
the gap.

Having seen the essential difference between the gapped and gapless phases, we
now explore more difference between the two gapped phases. From the different
values of Hall conductance, we have already seen that the difference between
phase-I and phase-III is due to their different topological properties. In
fact, explicit calculation of edged systems (not shown here) shows the
existence of gapless edge states in phase-I, while they are absent in
phase-III. More understanding can be obtained by studying their low energy
behavior. In phase-I, the low energy excitation can be described by the system
Hamiltonian expanded around the points $\mathbf{K}\mathtt{\equiv}(\frac{\pi
}{2},\pm\frac{\pi}{2}),H_{\mathbf{K}}$=$ I\mathtt{\otimes}%
(2k_{y}\sigma_{x}\mathtt{+}2k_{x}\sigma_{y}\mathtt{+}4\gamma_{1}\sigma
_{z})\mathtt{+}\gamma_{2}\sigma_{x}\mathtt{\otimes} I$ with energy
spectrum $E_{\mathbf{k}}\mathtt{=\pm}\gamma_{2}\mathtt{\pm}2\sqrt
{k^{2}\mathtt{+}4\gamma_{1}^{2}}$ and $\mathbf{k}$ the momentum
relative to $\mathbf{K}$. In this case, the Hamiltonian is of the
Dirac type with first-order differential operators. When the system
is in the phase-III, the low energy excitation can be described by
the system Hamiltonian expanded around the points $\mathbf{\bar{K}}$
=$(0,\pi)$ $[$or $(0,0)]$,
$H_{\mathbf{\bar{K}}}\mathtt{=} I\mathtt{\otimes}(2(1\mathtt{-}%
k_{y}^{2}/2)\sigma_{x}\mathtt{+}2(1\mathtt{-}k_{x}^{2}/2)\sigma_{y}%
)\mathtt{+}\gamma_{2}\sigma_{x}\mathtt{\otimes} I$ with energy
spectrum $E_{\mathbf{\bar{K}}}\mathtt{=\pm}\gamma_{2}\mathtt{\pm}%
2\sqrt{1\mathtt{-}k^{2}}$. Clearly, unlike phase-I, the Hamiltonian for
phase-III is a Schr\"{o}dinger type of second-order differential operators.
Also, since the $\gamma_{1}$ term disappears in $H_{\mathbf{\bar{K}}}$, the
low energy physics in phase III is thus insensitive to the vortex profile and
looks more like ordinary BI. These discussions reveal the essential difference
between TI (phase-I) and BI (phase-III).

Now we briefly discuss the case with quarter filling. In this case, as shown
in last section there are two phases: gapless critcal phase-I and gapped
phase-II. In phase-I the Hall conductance is not quantized (as shown in Fig.
3) and there is no well-defined localized charge. While in phase-II, the Hall
conductance is quantized at $1e^{2}/h$. The charge for each layer can be
obtained as $\delta Q$=$\eta\sigma_{H}e/2$=$1/2\mathtt{\times}1/2e\mathtt{=}%
e/4$, which is verified by our numerical calculation (not show here). The
system at quarter filling with fractional charge $e/4$ is an interesting
example obviously different from many other situations with fractional charge
$e$/2.

%\section{DISCUSSION}
Before ending our discussion, we address briefly the accessability of the
parameters. There are three important length scales in our bilayer system:
magnetic length $l_{B}$, the penetration length of the type-$\Pi$
superconductor $\lambda_{L}$, and the interlayer distance $d$. $l_{B}%
\mathtt{=}\sqrt{\frac{\hbar}{eB}}\mathtt{\simeq}\frac{25.6}{\sqrt
{B(\text{Tesla})}}$ nm$,$ so $l_{B}$ has several nanometers when the magnetic
field of $\mathtt{\sim}10$ T. The pinned Abrikosov lattice constant $a$ has
the same scale of $l_{B}$, which makes it easier to realize $\Phi_{0}/2$ per
plaquette than the natural solid with lattice constant of $\mathtt{\sim}$0.1
nm, requiring the magnetic field in order of $10^{4}$ T$.$ As discussed in the
Ref. [8], the vortices are well separated and the defects are well localized
from the estimation of $\lambda_{L}$. The interlayer distance $d$ can be tuned
in experiment and is related to the key parameter $t_{2}$ (or $\gamma_{2}$) in
our bilayer system. Unlike other natural systems such as bilayer graphene with
$\gamma_{2}\mathtt{<<}1$ \cite{eleven}, $\gamma_{2}$ can have values in a very
large regime by tuning the width of the wells (which form 2DEG) in the growth
direction, as well as the width and height of the barrier between the wells.

%\section{\qquad CONCLUSION}
\textbf{Conclusion} we have investigated the bilayer 2DEG adjacent to type-II
superconductor thin film. We find that the system undergoes quantum phase
transitions between several different phases through modulating the interlayer
tunneling: gapped phase/topological insulator, gapless critical phase/metal,
and gapped phase/band insulator. Depending on different (topological)
properties of each phase, the system shows different Hall conductance,
accompanying with appearance/disappearance of topological excitation with
fractional charge of $e/2$ or $e/4$.

%\begin{acknowledgments}
\textbf{Acknowledgments} This work was partially supported by NSFC
under Grants No. 10604010, No. 10874020, and No. 10574150, by CAEP
under Grant No. 2008B0102004, and by the National Basic Research
Program of China (973 Program) under Grants No. 2009CB929103 and No.
2006CB921300.
%\end{acknowledgments}


\begin{thebibliography}{99}                                                                                               %


\bibitem {one}R. Jackiw and C. Rebbi, Phys. Rev. D \textbf{13}, 3398 (1976)

\bibitem {two}W. P. Su, J. R. Schrieffer, and A. J. Heeger, Phys. Rev. Lett.
\textbf{42}, 1698 (1979).

\bibitem {wilczek}J. Goldstone and F. Wilczek, Phys. Rev. Lett. \textbf{47},
986 (1981).

\bibitem {three}D. C. Tsui, H. L. Stormer, and A. C. Gossard, Phys. Rev. Lett.
\textbf{48}, 1559 (1982).

\bibitem {three1}R. B. Laughlin, Phys. Rev. Lett. \textbf{50}, 1395 (1983).

\bibitem {four}C.-Y. Hou, C. Chamon, and C. Mudry, Phys. Rev. Lett.
\textbf{98}, 186809 (2007).

\bibitem {five}B. Seradjeh, C. Weeks, and M. Franz Phy. Rev. B \textbf{77},
033104 (2008)

\bibitem {six}C. Weeks, G. Rosenberg, B. Seradjeh, and M. Franz, Nat. Phys.
\textbf{3}, 796 (2007).

\bibitem {six1}G. Rosenberg, B. Seradjeh, C. Weeks, and M. Franz, Phy. Rev. B
\textbf{79}, 205102 (2009).

\bibitem {seven}D. J. Thouless, M. Kohmoto, M. P. Nightingale, and M. den
Nijs, Phts. Rev. Lett. \textbf{49}, 405 (1982).

\bibitem {note1}For most situation, the hopping amplitude between
nearest-neighbor sites is much larger than that between next-nearest-neighbor
sites, i.e., $\gamma_{1}<<1$ and $4\gamma_{1}<2\sqrt{2}$.

\bibitem {nine}S. Sachdev, Quantum phase transitions (Cambridge University
Press, Cambridge, England, 1999)

\bibitem {eight}P. Zanardi and N. Paunkovic, Phys. Rev. E \textbf{74}, 031123 (2006).

%\bibitem{ten} W. P. Su, J. R. Schrieffer, and A. J. Heeger, Phys. Rev. B
%\textbf{22}, 2099 (1980).


%\bibitem{franz08}B. Seradjeh, H. Weber, and M. Franz, Phys. Rev.
%Lett. \textbf{101}, 246404 (2008).


\bibitem {eleven}R. Dillenschneider and J. H. Han, Phys. Rev. B \textbf{78},
045401 (2008); Johan. Nilsson, A. H. Castro Neto, F. Guinea, and N. M. R.
Peres, Phys. Rev. B \textbf{78}, 045405 (2008).
\end{thebibliography}
\end{document}